\documentclass[5p,number, sort&compress, final]{elsarticle}
\usepackage{amsfonts}
\usepackage{amssymb}
\usepackage{graphicx}
\usepackage{times}
\usepackage{subfig}
\usepackage{hyperref}
\usepackage{lineno}

\journal{Astroparticle Physics}

\hypersetup{
bookmarks=false,             
unicode=true,                
pdftoolbar=true,             
pdfmenubar=true,             
pdffitwindow=false,          
pdfstartview={FitH},         
pdftitle={Using spherical wavelets to search for magnetically-induced alignment in the arrival directions of ultra-high energy cosmic rays},       
pdfauthor={M. Zimbres, R. Alves Batista, E. Kemp},         
pdfsubject={Wavelets},        
pdfcreator={M. Zimbres, R. Alves Batista, E. Kemp},        
pdfproducer={M. Zimbres, R. Alves Batista, E. Kemp},       
pdfkeywords={Wavelets} {Multiplets}, 
pdfnewwindow=true,      
colorlinks=true,        
linkcolor=red,          
citecolor=green,        
filecolor=magenta,      
urlcolor=cyan           
}

\begin{document}
\begin{frontmatter}
\label{firstpage}

\title{Using spherical wavelets to search for magnetically-induced alignment in the arrival directions of ultra-high energy cosmic rays}
\author[unicamp]{M. Zimbres} 
\ead{mzimbres@ifi.unicamp.br}
\author[uhh]{R. Alves Batista}
\ead{rafael.alves.batista@desy.de}
\author[unicamp]{E. Kemp}
\ead{kemp@ifi.unicamp.br}
\address[unicamp]{Instituto de F{\'{i}}sica ``Gleb Wataghin'' - Universidade Estadual de Campinas, 13083-859, Campinas-SP, Brazil}
\address[uhh]{II. Institut f\"ur Theoretische Physik - Universit\"at Hamburg, Luruper Chaussee 149, D-22761, Hamburg, Germany}


\begin{abstract}
Due to the action of the intervening cosmic magnetic fields, ultra-high energy
cosmic rays (UHECRs) {\color{black}can be deflected in such a way as to create} clustered
energy-ordered filamentary structures in the arrival direction of these
particles, the so-called multiplets. In this work we propose a new method
based on the spherical wavelet transform to identify multiplets in sky maps containing
arrival directions of UHECRs. {\color{black}The method is illustrated in simulations} with a multiplet
embedded in isotropic backgrounds with different numbers of events. The efficiency of the 
algorithm is assessed through the calculation of Type I and II errors.
\end{abstract}

\begin{keyword}
 spherical wavelets \sep ultra-high energy cosmic rays\sep cosmic magnetic fields \sep multiplets
\end{keyword}

\end{frontmatter}


\section{Introduction}

Cosmic rays were discovered more than one century ago. One remarkable feature 
of the cosmic ray spectrum is that it spans more than ten orders of magnitude, up
to hundreds of EeV (1 EeV $= 10^{18}$ eV). The spectrum roughly follows
an inverse power law, which means that the expected flux of particles at the highest
energies is extremely low compared to the lower energies. In fact, at energies of a 
few EeV,  only one particle per 
square kilometer per year is expected. Particles with energies $\gtrsim$ 1 EeV 
{\color{black}are referred to as ultra-high energy cosmic rays}.  Some experiments, 
such as the Pierre Auger Observatory and the {\color{black}Telescope Array Project}, have been designed to 
increase the statistics of events in this energy range. 
{\color{black} Despite the improved statistics, questions pertaining to the origin, nature 
and mechanisms of acceleration of these particles, remain unanswered}.

Due to the presence of galactic and extragalactic magnetic fields, charged
cosmic rays are expected to be deflected. Hence, incoming directions, as
measured by a detector, do not point back to the exact position of the
source. The magnitude of the deflection depends on the strength of the
intervening fields. In the case of charged particles the deflections are roughly
inversely proportional to the particle energy.  Therefore, for coherent fields,
the different Larmor radii described by cosmic rays can create filamentary
structures ordered by energy, {\color{black}known as multiplets}. This allows the reconstruction of
the source position, and consequently enhances the possibility to do astronomy with
UHECRs.

In this paper we propose a new method of identifying multiplets, based on the
spherical wavelet transform. The paper is organized as follows: in section
\ref{physics}, we review the physics underlying multiplets; in section
\ref{sec::swt} we present the wavelet transform from a pattern matching
algorithm point of view, and give motivations for its use in cosmic ray
physics; in section \ref{sec::theory} we present a novel algorithm to identify 
filamentary structures in cosmic ray maps; in section \ref{sec::analysis} the method
is applied to simulated data sets; 
in section \ref{sec::results}, we present our results; in section \ref{sec::conclusion} we
make our final remarks.

\section{Cosmic Magnetic Fields and Multiplets} \label{physics}

{\color{black} Several results  show that at ultra-high energies the cosmic ray spectrum contains
atomic nuclei}. Data from the High Resolutions Fly's Eye Experiment (HiRes) indicate
that UHECRs are probably  protons \citep{hires10}, whereas the results from the
Pierre Auger Observatory indicate that the composition tends to heavy nuclei at
the highest energies\footnote{Notice that this results strongly depends upon the
hadronic interaction model taken into account.} \citep{auger10}.  Despite this
controversy, one can consider that {\color{black}UHECRs are predominantly charged particles and, as such,
can be deflected by magnetic fields}.

The deflection expected for a UHECR of charge $Z$ due to the regular component
of the galactic magnetic field (GMF) is given approximately by \citep{harari02}:
\begin{equation}
	\label{eq:deflec}
	\delta \approx 53^\circ \frac{Z}{E} \left| \int\limits_0^L \frac{d\vec{r}}{\mbox{kpc}} \times \frac{\vec{B}}{\mbox{$\mu$G}} \right| {\ } \mbox{EeV},
	\label{deflection1}
\end{equation}
where $E$ is the energy of the particle, $\vec{B}$ the magnetic field, and $L$
the travel distance of the particle. Since $|\vec{B}|\sim \mu$G
and is coherent over lengths of $\sim$ 10 kpc, typical deflections are
$\sim$ 10$^\circ$ for protons of 10 EeV.

The expected deflection due to the turbulent
component is \citep{auger12}:
\begin{equation}
	\label{eq:deflec-turb}
	\delta_{turb}\approx 10^\circ \frac{Z{\ }{\mbox{EeV}}}{E} \frac{B_{rms}}{\mbox{$\mu$G}} \sqrt{\frac{L}{\mbox{kpc}}} \sqrt{\frac{L_c}{\mbox{50 pc}}},
	\label{deflection2}
\end{equation}
where $B_{rms}$ is the root mean square intensity of the magnetic field and $L_c$ is its coherence length.

\begin{figure}[h!]
\centering
\includegraphics[width=0.9\columnwidth]{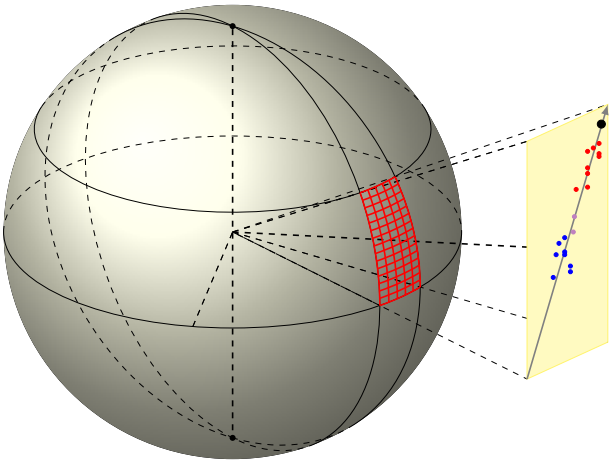}
\caption{Illustration of a multiplet on the tangent plane of a sky map. The dots represent events from higher (red) to lower (blue) energies. The black dot corresponds to the position of the source.}
\label{fig::sphere}
\end{figure}

According to equation (\ref{eq:deflec}) the deflection is inversely proportional
to the energy of the particle. Therefore it is possible that energy ordered
filamentary structures, the so-called multiplets, can be detected in cosmic ray maps
as shown in figure \ref{fig::sphere}. Another method to detect filamentary structures in the
arrival directions distributions of UHECRs was proposed by Harari {\it et al.}\citep{harari06}.

An interesting property of the multiplets
is that the position of the source can be reconstructed, allowing one to
identify UHECRs sources. A method to reconstruct the source position
of a multiplet was presented by Golup {\it et al.} \citep{golup09} and was used by the Pierre
Auger Collaboration to estimate the position of the sources for some possible
multiplet candidates \citep{auger12}.
In the aforementioned work, {\color{black} no statistically significant evidence 
for multiplets arising from magnetic deflections was found for energies above 20 EeV}.

It is important to notice that for some models of the galactic
magnetic fields such as the ones proposed by \citep{harari99,sun08,pshirkov11}
cosmic ray multiplets can be formed,  whereas in other models such as the one recently
proposed by \citep{jansson12} they are less likely to occur, due to the
strength of the field, especially the turbulent component.

The role played by extragalactic magnetic fields in the deflection of UHECRs is
not fully understood. Simulations of the propagation of UHE particles in the
large scale structure of the universe have been performed by several groups
\citep{sigl03,dolag05,armengaud05,das08}. However, these results are
contradictory and one cannot obtain a clear picture of the effects of the
extragalactic magnetic field for the deflection of UHECRs. 

\section{Wavelets on the sphere} \label{sec::swt}

{\color{black}In many branches of Physics, particularly Astrophysics} and Cosmology, wavelets
have been successfully applied to solve various problems, particularly related
to detection of signals. Wavelets on the plane have been widely used to denoise
cosmic microwave background  (CMB) maps \citep{cayon99,sanz99,gonzalez-nuevo06}.
However, the problem of identifying anisotropies in the distribution of arrival
directions of UHECRs has not been properly addressed, and only a few works
\citep{fay11,alvesbatista10,alvesbatista11,ivanov03a,ivanov03b} on this topic are
available in the literature.

{\color{black} Wavelets are commonly used in one and two dimensional data
analysis, but in recent years the interest in data lying on the sphere has
increased. This is due to experiments such as the Cosmic Background Explorer
(CoBE), the Wilkinson Microwave Anisotropy Probe (WMAP), the Planck Satellite
and the Pierre Auger Observatory, which make use of these kinds of data.
The need to process these data sets drove the interest on
new techniques, such as wavelets, under active development. Many interesting applications
of spherical wavelets can be found in \citep{martinez-gonzalez02,mcewen06a,mcewen06b,mcewen08a, mcewen07,mcewen08b}.
}

{\color{black} Wavelets can be particularly useful for cosmic ray data analysis, where we usually 
have to deal with a non uniform exposure. They are also useful to search for local structures in the sky,
with a defined position, such as point sources, and possibly an orientation, such 
as multiplets.}  An event by event analysis  is not viable, and the analysis in
harmonic space would be even harder since all local properties are lost.
Spherical wavelets come up as a good alternative to address these problems.
Other attractive features of wavelet analysis are:
\begin{itemize}
   \item any function can be exactly represented by its wavelet coefficients;
   \item local features of the signal can be {\color{black}enhanced in the wavelet domain},
   meaning that the number of coefficients needed to represent a given signal
   is reduced;
   \item it provides scale decomposition, making it possible to identify
   structures with different angular sizes and focus on the
   {\color{black} resolution of interest} in the signal;
   \item it does not rely on any tangent plane approximation, in the case of
   wavelets on the sphere.
\end{itemize}

Data representation in wavelet domain can be thought as something between pixel
and harmonic representation. Sometimes it is very convenient to decompose the
data to enhance properties that are not
clear in harmonic or pixel domain. Wavelets can be interpreted as local
analysis functions which can be rotated and/or dilated, to obtain
information regarding the signal morphology.

\subsection{Pattern matching on the sphere} \label{sec::matching}

In this subsection we show that the problem of finding a multiplet, or any
other pattern defined on the sphere, can be treated by the fast 
rotational matching algorithm \citep{kovacs02}.

Let  $f(\theta,\varphi)$ and  $h(\theta,\varphi)$ be two functions defined on
the sphere. Assume that $h$ is a rotated version of $f$, such that
$f=\Lambda(\alpha,\beta,\gamma)h$, with $\alpha$, $\beta$ and $\gamma$ being Euler
angles, and $\Lambda$ denoting the rotation operator in $SO(3)$.  

If we know that a rotated version of the pattern $f$ is present in $h$,
we can find its latitude, longitude and orientation on the sphere by
correlating all rotated versions of $f$ with $h$, and selecting the rotation
which maximizes the correlation
\begin{eqnarray}
C = \int\limits_{\mathbb{S}^2}h(\theta,\varphi)\overline{\Lambda f(\theta,\varphi)}d(\cos\theta)d\varphi. 
\label{corr}
\end{eqnarray}
We can parametrize the
rotations in terms of Euler angles. In this case the correlation function can
be denoted by $C = C(\alpha,\beta,\gamma)$. In other words, we want to
find the angles $\alpha$, $\beta$ and $\gamma$ for which $C$ is maximum.

The straightforward evaluation of $C$ is very time-consuming and not affordable depending
on the precision we are using to describe our signal.  Denoting the
band-limit of the signal by $B$, the complexity of equation (\ref{corr})
is $O(B^5)$. This complexity can be reduced to $O(B^4)$ by calculating the
correlation in the harmonic domain. So, we can write the spherical harmonic expansions
of $f$ and $h$ as
\begin{eqnarray}
f(\theta,\varphi) = \sum_{l=0}^{B-1}\sum_{m=-l}^{l}a_{lm}
Y^l_{m}(\theta,\varphi)  \label{eq::sph}\\
h(\theta,\varphi) = \sum_{l=0}^{B-1}\sum_{m=-l}^{l}b_{lm}
Y^l_{m}(\theta,\varphi),
\end{eqnarray}
with $Y^l_{m}$ being spherical harmonics. Using these expansions we can show
(see ref. \citep{kostelec08} for more details) that the correlation function
(\ref{corr}) can be cast as
\begin{eqnarray}
C &=& \sum_{l=0}^{B-1}\sum_{m,n=-l}^{l} \overline{a_{lm}}b_{ln} D^l_{mn}(\alpha,\beta,\gamma) \\ 
  &=& \sum_{l=0}^{B-1}\sum_{m,n=-l}^{l} \overline{a_{lm}}b_{ln} e^{im\alpha}d^l_{mn}(\beta)e^{in\gamma} ,
  \label{CC}
\end{eqnarray}
where $D^l_{mn}$ denotes the Wigner-D functions and $d^l_{mn}$ the small
Wigner-d functions.  This algorithm is called {\it Fast Rotational Matching},
abbreviated to FRM. It uses the Fast Fourier Transform (FFT) on both $\alpha$
and $\gamma$ to calculate equation \ref{CC}.  Depending on the tessellation
chosen, the FFT can be extended to the $\beta$ coordinate \citep{kovacs02}.

This approach to find a multiplet may not be an option because we do not
know exactly the pattern $f(\theta,\varphi)$ that describes the multiplet.
Moreover, for angular resolutions corresponding to a band limit higher than $B = 128$, 
too much memory would be required. For $B = 256$, for example,
approximately 4.5 GB of RAM would be necessary in our implementation.

In the next subsection we show that the difficulties mentioned above can be 
overcome by using a special family of directional wavelets instead of the pattern itself.

\begin{figure}[h!]
   \centering
   \includegraphics[scale=0.99]{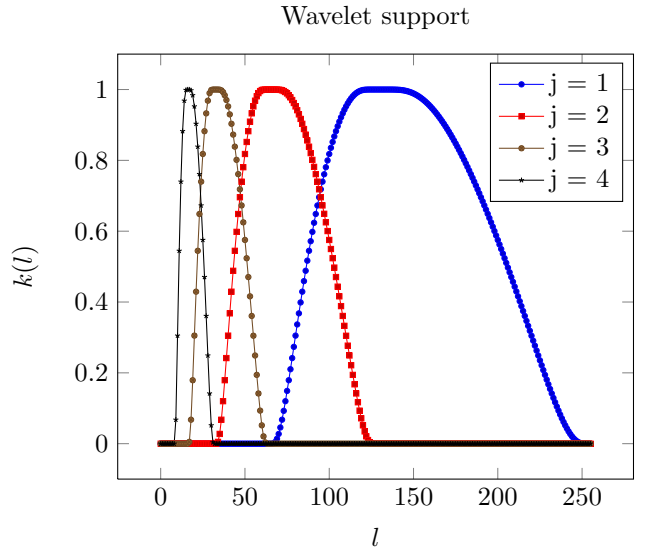}
   \caption{The figure shows the harmonic support in the frequency domain of the family of wavelets used, for $J = 8$.}
   \label{fig::kernel}
\end{figure}

\subsection{Pattern matching with directional wavelets}

When the function $f$ from equation \ref{eq::sph} is a wavelet, 
and $h$ is the signal of interest,
equation \ref{corr} is the definition of the forward spherical wavelet
transform, and the correlation function can be interpreted as the wavelet
representation of the signal $h$.

The wavelets used in this work are defined in the harmonic domain\footnote{For
the explicit definition and derivation of the mother-wavelet equation, refer to
ref. \cite{wiaux08}.}. Their harmonic representation has the
special property of allowing them to be split into a kernel and a directional
part
\begin{eqnarray}
b_{lm} = k(l)S_{lm},
\end{eqnarray}
where the kernel $k(l)$ is responsible for dilations and $S_{lm}$ is responsible for
 the directional properties of the wavelets. This split ensures that dilations do not affect directional properties, so that these two parts can be treated independently. 

\begin{table}
\centering
\caption{First column: scale $j$. Second column: wavelet support. Third
column: angular sizes in degrees for which the wavelet is sensitive.
Fourth column: maximum precision on the angular variables in degrees.}
\begin{tabular}{cccc}
 \hline
   $j$ & Support     & Angular size ($^\circ$)   & Precision ($^\circ$)  \\ \hline
   $0$ & $(256,128)$ & $(0.7,1.4)$    & $0.7$           \\ 
   $1$ & $(256,64)$  & $(0.7,2.8)$    & $0.7$           \\ 
   $2$ & $(128,32)$  & $(1.4,5.6)$    & $1.4$           \\ 
   $3$ & $(64,16)$   & $(2.8,11.3)$   & $2.8$           \\ 
   $4$ & $(32,8)$    & $(5.6,22.5)$   & $5.6$           \\ 
   $5$ & $(16,4)$    & $(11.3,45.0)$  & $11.3$          \\ 
   $6$ & $(8,2)$     & $(22.5,90.0)$  & $22.5$          \\ 
   $7$ & $(4,1)$     & $(45.0,180.0)$ & $45.0$          \\ 
   $8$ & $(2,1)$     & $(90.0,180.0)$ & $90.0$          \\ 
\end{tabular}
\label{table}
\end{table}

The kernel $k(l)$ at each scale $j$ has zero values at frequencies outside
the range $(2^{J-1-j},2^{J+1-j})$, {\color{black} where $J$ is the total number of scales}.
This interval is usually referred to as the
wavelet support. The support is related to the frequencies to which the wavelets
are sensitive. For a graphical representation see figure \ref{fig::kernel}. It is interesting
to notice in this figure that an adequate choice of $j$ can suppress some values of
$l$, making this method powerful even when we consider the exposure of a detector, which is 
usually {\color{black}associated with low values} of $l$.

In our implementation, for a given band limit $B$, the number of scales $J$ in the wavelet analysis
is given by $J = \log_{2}B$. Since we have used $B = 256$, then  $J = 8$.
For a physical interpretation, however, it is simpler to think in terms of
angular sizes. For that, we can convert the frequency range into angular
sizes using the formula $360^{\circ}/2l$, where $l$ is the frequency we are
interested in. In table \ref{table} the angular size to which each scale is 
sensitive is shown.  

\begin{figure}[h!]
   \centering
   \subfloat[$N = 1$]{
     \includegraphics[scale=0.57]{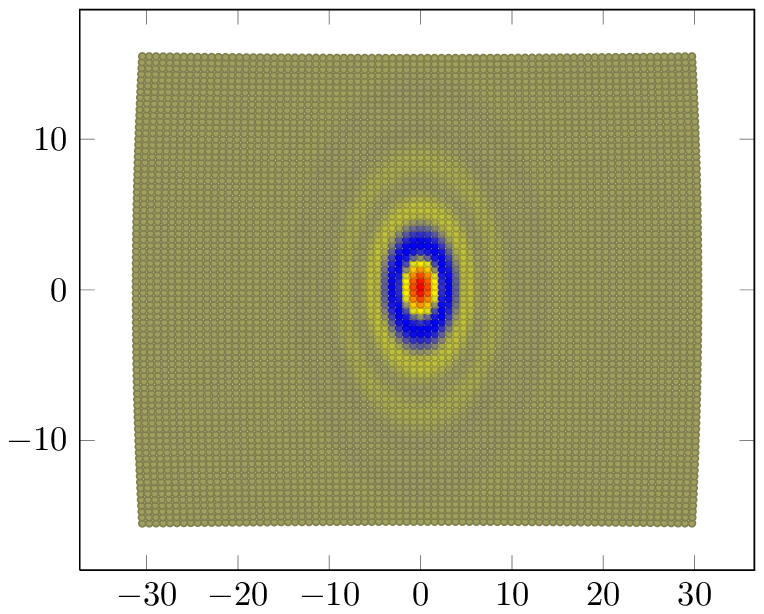}
   }
   \subfloat[$N = 127$]{
     \includegraphics[scale=0.57]{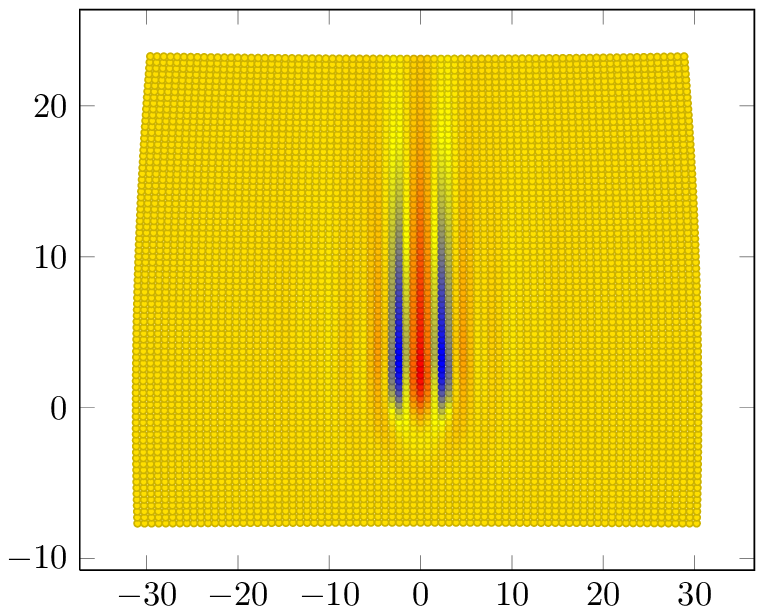}
   }
   \caption{Comparison of wavelet with parameters $(J,j,N) = (8,2,1)$ and $(J,j,N) = (8,2,127)$.
   Both axes are in units of degrees. The wavelet shown in figure (b) is the one used in the analysis.}
   \label{fig::wav-show}
\end{figure}

The sensitivity of the wavelet in finding the angle $\gamma$ can be controlled
by imposing a band limit on $S_{lm}$. Denoting this band limit by $N$, then 
$S_{lm} = 0$ for all $m \geq N$. If the directional features of the signal are not relevant for
the analysis, low values of $N$, such as $N=1$, can be used. The maximum value of $N$
at scale $j$ is given by $N = 2^{J-j+1}$.  The precision of the orientation
 for a given $N$ is given by $\Delta\gamma =
180/N$. In this work, we have used $N = 127$, shown in figure \ref{fig::wav-show}, which gives a precision
$\Delta\gamma = 1.42$ degrees. This value is a good compromise between computational resources
and the required precision.

\section{Looking for multiplets} \label{sec::theory}

We have developed an algorithm that uses spherical wavelets to identify and locate 
filaments in maps containing arrival directions of UHECRs. The algorithm is described below.

\begin{enumerate}
   \item From a set of events, calculate the function $h(\theta,\varphi)$ 
   that represents the signal. This involves choosing a tessellation for the sphere,
   and counting the total number of events contained in each pixel in the sky.

   \item Calculate the Fourier expansion of $h(\theta,\varphi)$, resulting 
   in the coefficients $a_{lm}$ (see equation \ref{eq::sph}).

   \item Choose the appropriate wavelet. For the family of wavelets used,
   this involves choosing three parameters, the maximum scale $J$, 
   the scale at which the analysis is performed $j$ and the azimuthal band limit,
   controlled by the parameter $N$. Ideally, the angular size of the wavelet will
   match the size of the multiplet. This choice relies on \ref{table}. 

   \item Calculate $C(\alpha,\beta,\gamma)$ from equation (\ref{CC}).
 
   \item Select all $(\alpha,\beta,\gamma)$ such that
   $|C(\alpha,\beta,\gamma)| > C_0$, where $C_0$, is threshold value.

   \item Select the events at each location
   $(\alpha,\beta,\gamma)$ (see section \ref{sec::selecting}).
   This will result in $m$ groups of events.

   \item Discard all groups of events for which $n < n_0$, where $n$ is the number of
   events in the group, and $n_0$ is a threshold value.

   \item For each group of events, calculate the correlation $c$ of the graph 
   $\delta\times 1/E$ (see equation \ref{corr_corr}).

   \item Accept as a multiplet candidate all groups for which $|c| > c_0$,
   where $c_0$ is a threshold value.
   (see section \ref{sec::selecting})
\end{enumerate}

\subsection{Establishing thresholds} \label{sec::thresholds}

To carry out all steps of this algorithm one needs to establish three threshold
values: the wavelet threshold $C_0$ used in step $5$, the minimum number of
events $n_0$, used in step $7$, and the minimum correlation $c_0$, used in step
$9$. 

To calculate the thresholds, we have used simulated sky maps containing arrival
directions of events isotropically distributed. No magnetic fields are considered in 
this case, so that if a multiplet is identified by our method, it certainly happened 
by chance. This allows us to establish the threshold values $C_0$, $c_0$, and
$n_0$ {\color{black}by using the algorithm previously presented in section} \ref{sec::theory}. 

{\color{black}To estimate $c_0$ we use M simulations of isotropic skies, where
each simulated sky follows the same injection spectrum and exposure as the data
that is being analysed}. $C_i$ is
the largest wavelet coefficient obtained for each realization $i$, according to
step $5$ of the algorithm.

By selecting only one value of $C$ for each realization, we have a single 
correlation coefficient $c_i$ for each of the simulated isotropic skies, 
calculated at step $8$ of the  algorithm. 

From $M$ wavelet coefficients $C_i$ and correlations $c_i$, we choose 
\begin{equation}
C_0 = \bar{C} + r_C\sigma_C \ \ \mbox{and} \ \ \ c_0 = \bar{c} + r_c\sigma_c,
\label{eq::thresholds}
\end{equation}
where $\bar{C}$ and $\bar{c}$ denote the average values over $M$ realizations, and
$\sigma_C$ and $\sigma_c$ their respective standard deviations. The numbers
$r_C$ and $r_c$ are chosen according to the error the user is willing to accept.
In this work we set $r_C=1$ and $r_c=1$.

The threshold value for the number of events in the multiplet, $n_0$,  is chosen 
depending on the specific problem. In this analysis, for example, we have chosen 
$n_0 = 10$, {\color{black}following ref}. \cite{auger12}.

\subsection{Selecting events} \label{sec::selecting}

\begin{figure}[h!]
   \begin{center}
   \includegraphics[width=0.8\columnwidth]{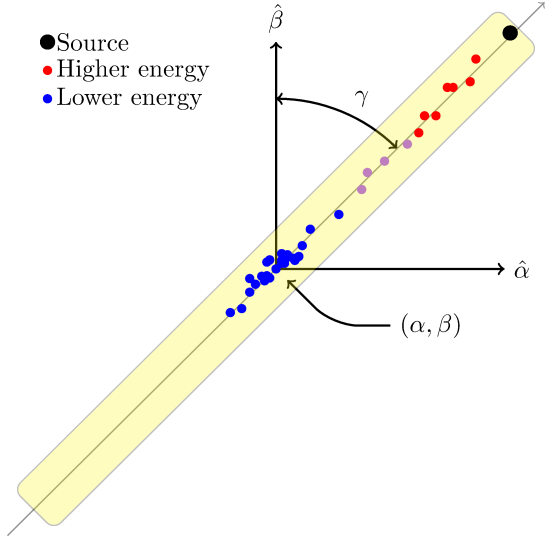}
   \caption{Illustration of a multiplet along a segment. The dots represent
   events from higher (red) to lower (blue) energies. The angles
   $(\alpha,\beta,\gamma)$ are the Euler angles associated with the multiplet.}
   \label{fig::plane}
   \end{center}
\end{figure}

To select events around the position where a wavelet has a 
high magnitude we need the Euler angles $\alpha$ and $\beta$, to 
provide a location in the sky, and the angle $\gamma$, to provide an
orientation. A multiplet can, in principle, describe an arbitrary curve on the 
celestial sphere, depending on the intervening magnetic fields.
To be general for all possible shapes we have used small segments as shown on
figure \ref{fig::plane}. Each segment should be small enough so that the curve
described by the events is approximately a straight line inside the segment.

A small deflection angle ($ \lesssim 10^{\circ}$) can be written as 
$\delta = \Delta s/r$, which in spherical coordinates takes the form
\begin{equation}
\Delta s^2 =  r^2((\Delta\beta)^2 + \sin^2\beta (\Delta\alpha)^2),
\label{displacement}
\end{equation}
with $\Delta s$ being the linear distance over the surface of a sphere of radius $r$.

Now we note that the unit vectors $\hat{\alpha}$ and $\hat{\beta}$, which point in the 
direction where $\alpha$ and $\beta$ vary, are orthogonal and span the plane tangent to the sphere 
at $(\alpha,\beta)$. Using equation \ref{displacement} we can write a displacement vector, 
tangent to that point as
\begin{equation}
\vec{d} =  r(\Delta\beta)\hat{\beta} + r\sin\beta (\Delta\alpha)\hat{\alpha}.
\label{d}
\end{equation}
It is convenient to write $\vec{d}$ in terms of $\hat{L}$, aligned
with the wavelet, and $\hat{W}$, which is perpendicular to it.  These vectors are
related to $\hat{\alpha}$ and $\hat{\beta}$ by a rotation of an angle $\gamma$ 
in the tangent plane, as shown below:
\begin{eqnarray}
\hat{\alpha} & = & \cos\gamma\hat{L} - \sin\gamma\hat{W} \\
\hat{\beta} & = & \sin\gamma\hat{L} - \cos\gamma\hat{W}.
\end{eqnarray}
We can now cast equation \ref{d} as
\begin{eqnarray}
\vec{d} = d_{\hat{L}}\hat{L} + d_{\hat{W}}\hat{W},
\label{vec}
\end{eqnarray}
where
\begin{eqnarray}
\label{distance2}
d_{\hat{L}} & = & r(\Delta\beta\cos\gamma + \sin\beta\Delta\alpha\sin\gamma) \\
d_{\hat{W}} & = & r(-\Delta\beta\sin\gamma + \sin\beta\Delta\alpha\cos\gamma).
\label{distance1}
\end{eqnarray}

With equations \ref{distance2} and \ref{distance1}, we can select events in the
sky at the position of maximum wavelet coefficient. The only information required are the Euler angles
$\alpha$, $\beta$, and $\gamma$, since $r=1$ for the celestial sphere.

\section{Analysis} \label{sec::analysis}

To simulate both the background and the events belonging to the multiplet, we
have used the software CRPropa \citep{kampert13}.  Protons were injected
following a typical $E^{-2.2}$ spectrum, for $15<E<40$ EeV.  The effective
detector size is a sphere of {\color{black}diameter} 1 Mpc, and the source was assumed to be in
our local universe, at a distance of 30 Mpc from the detector. {\color{black} If
the events reach the sphere, a flag ``detected" is raised. Otherwise, the events
are rejected. With these settings we achieve an angular resolution of $\approx$
1.9$^\circ$.} 

For the isotropic data sets, the events were simulated in random positions of
the sky.  

We have divided our analysis in two parts. First we analyze an isotropic
distribution of events to establish the thresholds,  as explained in section
\ref{sec::thresholds}.
In the second part of the analysis we used a simulated multiplet {\color{black}
composed of 10 events}, embedded in the same isotropic datasets previously
used. {\color{black} We have chosen 10 events based on \citep{auger12}, where 10
is the lower number of events found on a multiplet.} {\color{black} The multiplet
was simulated in the same direction each time, since by spherical symmetry the
simulation is unchanged if both the source position and the magnetic field are
rotated by the same angle. On the other hand, keeping the magnetic field
constant and varying the source position could be a source of confusion, since
the perpendicular component of the magnetic field would also vary, resulting in different
multiplets each time. In the special case where event trajectories are
parallel to the magnetic field, the analysis is pointless,
since no multiplet is formed.}

\begin{figure}[h!]
   \begin{center}
   \includegraphics[width=\columnwidth]{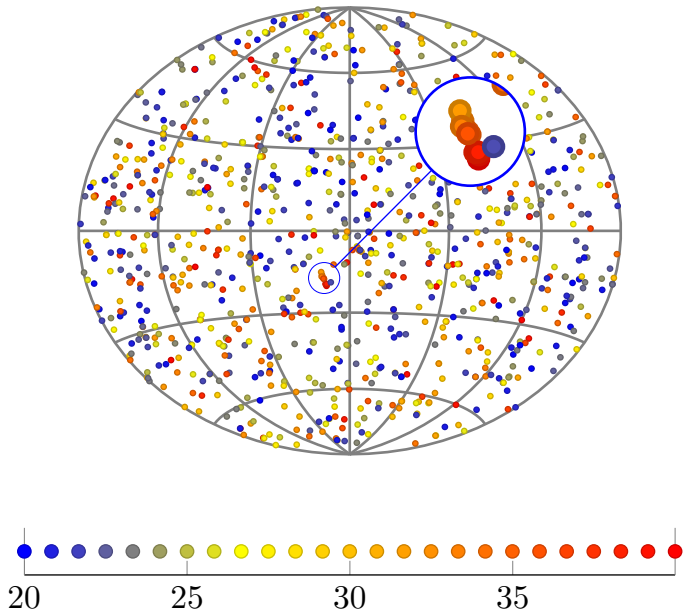}
   \caption{Simulated multiplet containing 10 events embedded in a background
   of 1000 events. The color scale corresponds to the energy of the events in
   units of EeV. The zoom shows more closely the events of the simulated
   multiplet.}
   \label{fig::sim-mult}
   \end{center}
\end{figure}

The multiplet formation was induced by an uniform magnetic field of 1 nG
oriented in the $\hat{z}$ direction.
Nevertheless, the results are totally independent of the method used to generate
the multiplets, since it depends only on the position of the events, and on the energy-
deflection correlation.
The number of isotropic events range between 100 and 1000 events, in steps of 100, and a
total of 1000 realizations for each of these values. The multiplet embedded in
one of the isotropic skies containing 1000 isotropically distributed events can
be seen in figure \ref{fig::sim-mult}. 

The simulated multiplet has a correlation coefficient $c = 0.99$ when no background is present.
$c$ corresponds to the Pearson's coefficient of the $\delta \times E^{-1}$ graph, given by:
\begin{equation}
	c=\frac{\sum\limits_{i=1}^{n} \left(  \frac{1}{E_i} - 
	\left<\frac{1}{E}\right> \right) \sum\limits_{i=1}^{n} \left(  \delta_i - 
	\left<\delta\right> \right) }{\sqrt{\sum\limits_{i=1}^{n} \left(  \frac{1}{E_i} - 
	\left<\frac{1}{E}\right> \right)^2} \sqrt{\sum\limits_{i=1}^{n} \left(  \delta_i - \left<\delta\right> \right)^2}}.
\label{corr_corr}
\end{equation}

The scale at which we performed the analysis was chosen based on table
\ref{table}. At scales $j = 2$ and $3$ the angular size of the wavelet is close
to what we would expect for a multiplet similar to the simulated one, i.e. 
$10^{\circ}\times 2^{\circ}$. Since at scale $j = 3$ the maximum precision 
of the angular variable is $2.8^\circ$, we have used $j = 2$ for the analysis.
The size of the segments are $10^{\circ}\times 2^{\circ}$, to match the dimensions of
the wavelet.

\section{Results and Discussion} \label{sec::results}

In this section we present the results of the proposed algorithm, when applied to the
simulations. 

\begin{figure}[h!]
   \begin{center}
   \includegraphics[width=0.95\columnwidth]{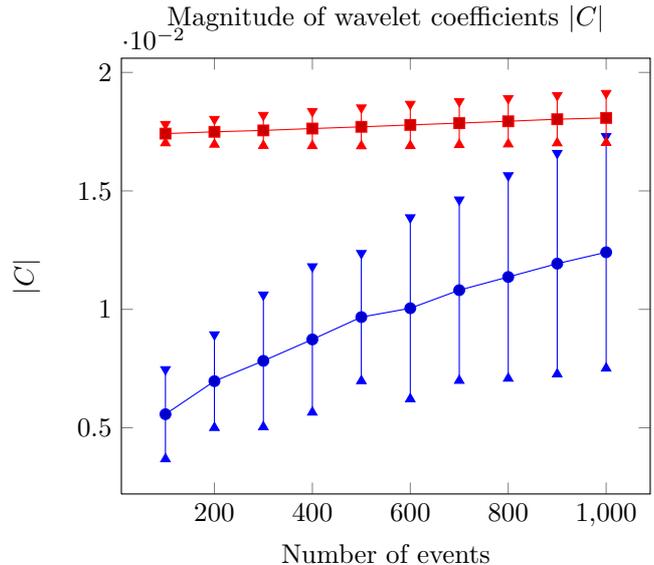}
   \caption{Magnitude of the largest wavelet coefficients in all the 1000 realizations, as a function of the number of events in the background. Blue circles correspond to the purely isotropic case, and red squares to the multiplet embedded in the isotropic background. The error bars are the standard deviation of the wavelet coefficients for all the realizations.}
   \label{fig::magnitude}
   \end{center}
\end{figure}

{\color{black}Since the magnitude of the wavelet coefficients are used} to select directions of interest
in sky maps, {\color{black}i.e. Euler angles}, establishing the wavelet coefficient threshold is a crucial
part of the analysis. The wavelets should have enough sensitivity to distinguish the cases
of interest (isotropic data sets with an embedded multiplet and without it). Figure \ref{fig::magnitude} 
shows the average values of the magnitude of the largest
wavelet coefficients found in each realization, for the two data sets. The error bars
are the standard deviation of the 1000 realizations of isotropic skies.
In this particular case, for a multiplet composed of 10 events, the error bars start overlapping
at around 1000 events, which corresponds to a fraction of events from the multiplet 
with respect to the background of $\approx 10^{-2}$.

\begin{figure}[h!]
   \begin{center}
   \includegraphics[width=0.95\columnwidth]{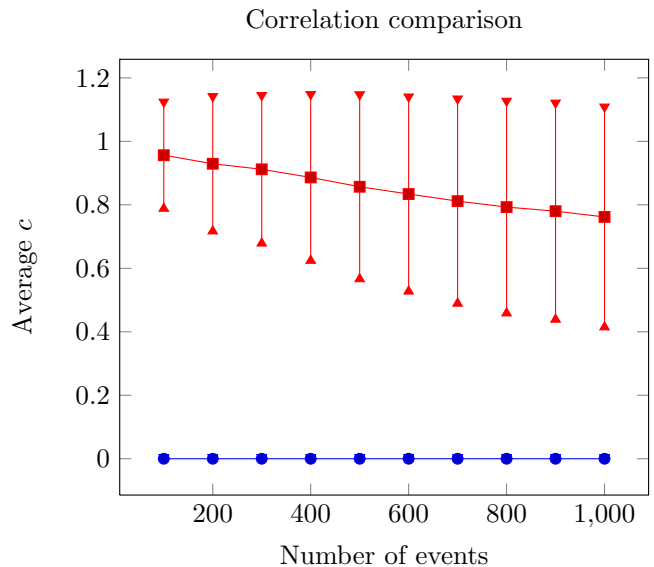}
   \caption{Correlation coefficients of the 1000 realizations, as a function of the number of events in the background. Blue circles correspond to the purely isotropic case, and red squares to the multiplet embedded in the isotropic background. The error bars are the standard deviation of the correlation coefficients for all the realizations.}
   \label{fig::correlation}
   \end{center}
\end{figure}

The correlation coefficient is the main observable to properly characterize a multiplet.
In this analysis we have used $n_0=10$, which means that a candidate will only be considered a multiplet if it is composed by, at least, 10 events. We also have to determine the value of $c_0$, which is the value of the threshold correlation coefficient. It is expected that the greater the number of isotropically distributed events, which corresponds to the background, the smaller the correlation coefficient. This is corroborated by figure \ref{fig::correlation}, 
which shows that the correlation coefficient monotonically decreases for larger number of background events, indicating a contamination
of the multiplet by the background. This allows us to establish a threshold value for the correlation coefficient.
For 1000 isotropic events the lowest value of $c_0$ considered is $c_0=0.4$, according to equation \ref{eq::thresholds}.

\begin{figure}[h!]
   \begin{center}
   \includegraphics[width=\columnwidth]{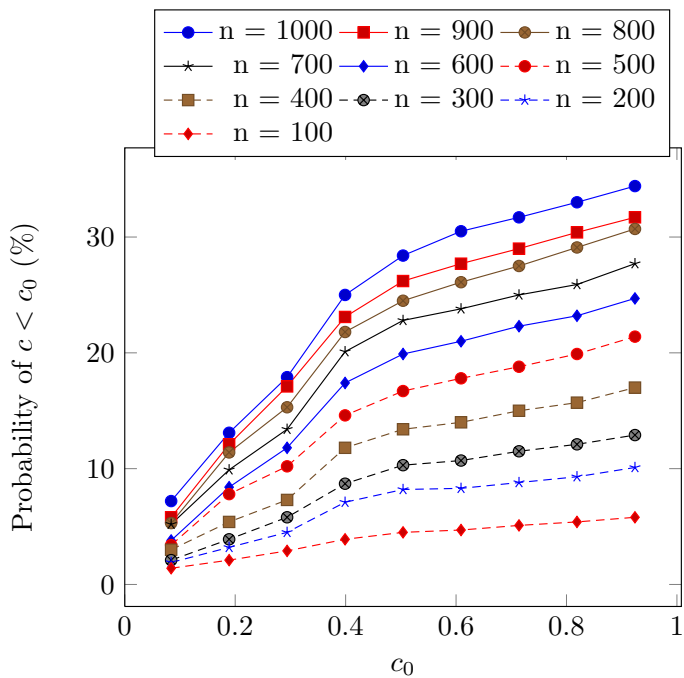}
   \caption{ Probability of having $c < c_0$ (Type I error) when the multiplet is known to be present in the data set. Each curve corresponds to a different fraction of events of the multiplet with respect to the isotropic background.}
   \label{fig::compare}
   \end{center}
\end{figure}

It is important to address the questions related to the probability of the algorithm finding 
a multiplet with $c>c_0$ in a given data set, when no multiplets are known to be present 
(Type II error), and the probability of a multiplet having $c<c_0$ if it is known to be present
in the data (Type I error).  In the isotropic simulations, no multiplets with $c>0.4$ were 
identified for a {\color{black}fraction of events equal to} $10^{-2}$, implying that the maximum 
Type II error of the method is $10^{-3}$. Figure \ref{fig::compare} shows the Type I error introduced by our method.
 It is worth mentioning that for the most conservative case, 
which has the lowest fraction of number of events of the multiplet with respect to the 
number of events of the background, the Type I error introduced by using a threshold 
value $c_0=0.4$ is approximately 25\%, whereas for the least conservative case it is 
around 2\%. For a high threshold value such as $c_0=0.9$, this error is of the order of 
30\% for the most conservative case. It is important to stress that the estimated Type I and II errors depend on the 
confidence interval adopted, set by the parameters $r_C$ and $r_c$.

\section{Conclusion}
\label{sec::conclusion}

We have developed a wavelet based analysis method to identify multiplets in maps 
containing arrival directions of UHECRs. We have illustrated the method by applying it
to a hypothetical scenario with a uniform magnetic field of 1 nG, considering several 
fractions of events from the multiplet with respect to the background, down to a fraction
of $10^{-2}$. All the parameters used in this study are rather arbitrary. These choices
were made to illustrate the method, and not to evaluate its overall performance.

The observables used to accept or reject the multiplet candidate were the magnitude $C$
of the wavelet coefficient at the multiplet location $(\alpha, \beta, \gamma)$, the correlation
$c$ of the deflection versus inverse of the energy graph and the total number
of events in the multiplet $n_0$. 

The efficiency of the method was assessed by comparing the analysis when applied to 
isotropically distributed events with and without an embedded multiplet. The probability of 
wrongly accepting a candidate multiplet with a correlation coefficient $c$ above the threshold
value $c_0 > 0.4$ is below $10^{-3}$. The probability of wrongly rejecting a candidate multiplet, when it
is known to be present in the data, is in the most conservative case, approximately 30\%. 
This value goes down to 3\% by decreasing the number of isotropic events in the data set. For the 
adopted correlation coefficient threshold, these values are, respectively, 25\% and 2\%.

Even though we have assumed a uniform detector exposure for the analysis previously presented, the results
also hold for a non uniform exposure, {\color{black}since the ideal parameters for a multiplet search} in 
the wavelet analysis {\color{black}suppress low frequency modes that could be related to} the coverage of
the cosmic ray detector. The only caveat is that, in this case, the comparison data sets should
follow an isotropic distribution modulated by the exposure of the detector.

The method is focused on the search of multiplets, but it can be adapted 
for generic application to related problems in astrophysics, particularly the ones 
involving the search of filamentary structures in sky maps.

All results showed in this paper can be easily reproduced with the software SWAT (Spherical 
Wavelet Analysis Tool), developed by the authors of this paper. In case of interest in the
code, please contact us.

\section*{Acknowledgments}

\noindent The authors wish to thank the Pierre Auger group of the University of Wuppertal (UW), 
where part of this work has been written, and part of the simulations done,
 and specially to Nils Nierstenh\"ofer for  the valuable help with the simulations. 
We would also like to thank Funda\c{c}\~ao de Amparo \`a Pesquisa do Estado de 
S\~ao Paulo (FAPESP) for the financial support through grants 2010/07359-6 and 2010/04743-0. 
Part of the simulations were performed in the computational system available at Instituto de
F\'isica de S\~ao Carlos, Universidade de S\~ao Paulo (IFSC-USP), granted by FAPESP (2008/04259-0).
This project was also supported by Coordena\c{c}\~ao de Aperfei\c{c}oamento de N\'ivel Superior (CAPES).

\label{lastpage}

\end{document}